\documentclass[10pt,letterpaper]{article}
\usepackage[T1]{fontenc}
\usepackage[utf8]{inputenc}
\usepackage{authblk}
\usepackage{color}
\usepackage{amsthm}

\usepackage{graphicx}

\usepackage{subfigure}
\usepackage{setspace}
\usepackage[top=1.0in, bottom=1.0in, left=1in, right=1in]{geometry}
\usepackage[british]{babel}
\usepackage[style=apa,backend=biber,sortcites=true,natbib=true]{biblatex}
\DeclareLanguageMapping{british}{british-apa}
\makeatletter

\providecommand{\tabularnewline}{\\}

\usepackage{mathtools}% Bonus

\providecommand{\definitionname}{Definition}
\providecommand{\theoremname}{Theorem}

\numberwithin{equation}{section}
\numberwithin{figure}{section}
\numberwithin{table}{section}
\theoremstyle{definition}
\newtheorem*{defn*}{\protect\definitionname}
\theoremstyle{plain}
\newtheorem{thm}{\protect\theoremname}

\begin{document}

\title{Bias Correction For Paid Search In Media Mix Modeling}
\author{Aiyou Chen, David Chan, Mike Perry, Yuxue Jin, \authorcr
  Yunting Sun, Yueqing Wang, Jim Koehler}
%\thanks{E-mail for correspondence: aiyouchen@google.com}
\affil{Google Inc.}
\date{Last Update: \today}
\maketitle

\begin{abstract}
  Evaluating the return on ad spend (ROAS), the causal effect of advertising on
  sales, is critical to advertisers for understanding the performance of their
  existing marketing strategy as well as how to improve and optimize it. Media
  Mix Modeling (MMM) has been used as a convenient analytical tool to address
  the problem using observational data. However it is well recognized that MMM
  suffers from various fundamental challenges: data collection, model
  specification and selection bias due to ad targeting, among others
  \citep{chan2017,wolfe2016}.

  In this paper, we study the challenge associated with measuring the impact of
  search ads in MMM, namely the selection bias due to ad targeting. Using
  causal diagrams of the search ad environment, we derive a statistically
  principled method for bias correction based on the \textit{back-door}
  criterion \citep{pearl2013causality}. We use case studies to show that the
  method provides promising results by comparison with results from randomized
  experiments. We also report a more complex case study where the advertiser
  had spent on more than a dozen media channels but
  results from a randomized experiment are not available. Both our theory and
  empirical studies suggest that in some common, practical scenarios, one may
  be able to obtain an approximately unbiased estimate of search ad ROAS.
\end{abstract}
\maketitle

\section{Introduction and problem description}

Evaluating the return on ad spend (ROAS) is a fundamental problem in marketing.
Many advertisers use multiple media channels to maximize their reach to
potential customers. Media mix modeling (MMM) is an analytical approach
(e.g. multivariate regression) first proposed by
\citep{borden1964,mccarthy1978} using observational data
(e.g. price, media spend, sales, economic factors) to estimate and forecast
the impact of various media mix strategies on sales.  While MMM has been
adopted by many Fortune 500 companies, various limitations have been
well-recognized, for example, data collection, selection bias, long-term
effects of advertising, seasonality and funnel effects, see
\citep{chan2017,wolfe2016} for discussion.

A typical MMM at a brand level can be described as a regression model
\citep{jin2017}, where the dependent variable is a key performance indicator (KPI),
often sales, and independent variables include various media inputs (e.g. spend levels, impressions or GRPs), product price, economic factors, competitors' marketing activities, etc, usually measured per market area on the daily, weekly or monthly basis. The value of the model to the advertiser is in the causal estimates of the set of media effects; causal inference is known to be notoriously hard with observational data \citep{imbens2015}. One of the major challenges to valid causal inference in MMM is selection bias due to ad targeting. Ad targeting is common across many different media channels, but is particularly acute in digital channels. Selection bias from ad targeting arises when an underlying interest or demand from the target population is driving both the ad spend and the sales.  See \citet{varian2016causal} for a formal mathematical description of selection bias.

In reality, advertisers often spend more when there is stronger
demand for their product. As a result, a naive regression which measures the change
in sales relative to the change in ad spend leads to over-estimates of
ROAS. A heuristic explanation is that the change in sales could be caused by a change in either consumer demand or ad spend or both, while the naive method ignores the change in consumer
demand. Evaluation of media effects from observational studies is questionable in general due to the risk of selection bias and related problems, see \citep{farahat2012effective,papadimitriou2011display,lewis2014online,lewis2011here,blake2015} and references therein.

In this paper, we study the selection bias issue in search ads in the context of media mix modeling. Using causal diagrams of the search ad environment, we derive a statistically principled method for paid {\bf s}earch {\bf b}ias {\bf c}orrection in MMM (SBC) based on the back-door criterion from the literature of causal
inference \citep{pearl2013causality}. We have carried out various case studies
using randomized experimental results as a source of truth, which show that SBC
provides promising results. Both our theory and empirical studies suggest that in some common, practical scenarios one may be able
to obtain approximately unbiased estimation for paid search ROAS without
solving all the challenges in MMM, such as funnel effects and selection
bias in non-search media channels.

The rest of the paper proceeds as follows: Section \ref{sec:relatedwork}
reviews related work; Section \ref{sec:prelim} describes the back-door
criterion; Section \ref{sec:Methodology} derives our SBC method and Section
\ref{sec:Implementation} describes the implementation procedure; some real case
studies are reported in Section \ref{sec:case123} in comparison with results
from randomized experiments, and in Section \ref{sec:case4} a more complex case
study is reported;\footnote{Disclaimer: All data analysis reported in this paper
was done with proprietary Google data
  and results may not be the same by using publicly available Google search
  data.} the conditions and limitations of the method are further discussed in
Section \ref{sec:discussion}.

\section{Related work} \label{sec:relatedwork}

There have been several research efforts focused on evaluating search ad effectiveness in the industry. Randomized experimentation is the gold standard.
%%. For example, Google Analytics uses randomized geo experiments
Some Google research has been reported in this direction \citep{vaver2011,vaver2012periodic,kerman2017tbr}. See \citep{blake2015} and \citep{farahat2012effective} for some examples of large-scale randomized experiments as well as comparison with non-experimental studies, carried out by eBay and Yahoo respectively. Due to practical limitations in implementing randomized experiments, the industry has been actively looking for alternative solutions based on observational studies, aside from media mix modeling. These can be summarized as follows.

The first type of research makes use of user-level data. The main idea is to compare users who were exposed to the ads with ones who were not exposed to the ads, either by propensity matching or covariate adjustment by regression. This type of methods are commonly employed in the industry but its risk is also well recognized, see examples in \citep{chan2010,lewis2011here,gordon2016}.

The second type of research makes use of aggregate data at a campaign level. The main idea is to estimate the difference in a KPI that a campaign may have made by comparing the observed KPI from the campaign with the counterfactual value had the campaign not happened. For example, researchers at Google \citep{hal2009adclick,chan2011incremental,brodersen2015,brodersen2017} have proposed various parametric models which use pre-campaign data to predict such counterfactual values.

The third type of research makes use of query-level data \citep{liu2012click}.
Liu assumed ad serving pseudo-randomness between organic search and paid search, and based on that derived an estimate of incremental value of ad impressions to ad clicks.

The first type of methods is less relevant to this study as MMM-related KPIs are
usually hard to collect at the user level. MMM data usually consist of various
campaigns across multiple media, which rule out direct application of the second
type of methods. Liu's work in the third type is closest to
ours in the spirit of looking into the search ad mechanism. His method
is based on query-level data. Our method works with aggregate data and does not assume randomness in ad serving.

There are other works on measuring search ad effectiveness, see for
example \citet{lysen2013} for measuring the incremental clicks impact of mobile search advertising, \citet{sapp2017} on near impressions, \citet{narayanan2015position} for measuring position effects
with regression discontinuity and \citet{rutz2011zooming} for using
both aggregate data and consumer level data.

\section{Preliminary to Pearl's causal theory}\label{sec:prelim}

A causal diagram is a directed acyclic graph (DAG), representing causal relationships between variables in a causal model. It comprises of a set of variables, represented as nodes of the graph, defined as being within the scope of the model. An arrow from node $i$ to another node $j$ represents
causal influence from $i$ to $j$, i.e. all other factors being equal, a change
in $i$ may cause changes in $j$. Below we first describe an example of causal diagram about search ad and then introduce the key concept of Pearl's causal theory that our estimation methodology will be based on.

\subsection{Causal diagram for search ads}\label{subsec:causal.diagram}

Consider a simplified causal diagram about how search ads affect sales value (e.g. sales revenue, or number of sales) based
on Google's search ad mechanism \citep{varian2009online} as follows.

Suppose that a user submits a search query (say ``flower delivery'') to www.google.com. There are typically two consequences: 1) the user would see a list of URLs plus a few lines of description in the main body of the search pages, called organic results, which are ranked by the search engine based on their relevance to the search query; 2) if the search query matches certain  keywords targeted by a set of advertisers, then the ads to be shown on the page will be chosen by auction. The auction considers various factors including bid, ad quality and advertiser homepage quality. The user may click on some URLs from organic results or click on the ads, and then land on some flower delivery websites to make an order.

\begin{figure}
\begin{centering}
\includegraphics[width=1.5in, height=2in]{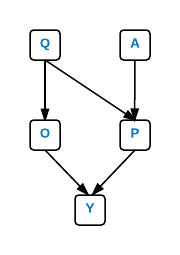}
\par\end{centering}

\caption{\label{fig:A-causal-diagram}A causal diagram for search ad at a query level, where $Q$ stands for the number of relevant queries, $A$ stands for auction factors, $O$ stands for organic search results, $P$ is the number of paid search impressions, and $Y$ stands for the sales value.}
\end{figure}

For this search event, let $A$ represent the auction factors, $Q$ be the search query controlled by a search user, $P$ indicate the presence of a paid search impression, and $O$ be organic search results. Given the query $Q$, $O$ is determined by the search engine\footnote{Per discussion with Hal Varian, personalized search is very limited and is only relevant for repeated searches. See https://googleblog.blogspot.com/2009/12/personalized-search-for-everyone.html.} and $P$ is determined by the search engine and other parties in the auction. Let $Y$ be the sales value. The causal path goes as follows: 1) $Q$ has two consequences $P$ and $O$; 2) $P$ is affected by both $Q$ and $A$; 3) $Y$ is affected by both $O$ and $P$. Therefore intervention on $P$ has direct effect
on $Y$, while intervention on $A$ does not have effect on $Y$
unless it causes changes in $P$.
%%Assume that other parties' actions and user information do not change (and thus can be omitted from the graph),
The causal diagram can be described by the directed acyclic graph shown in Figure \ref{fig:A-causal-diagram}.

Note that Figure \ref{fig:A-causal-diagram} makes an implicit assumption: Given a search query $Q$, organic search content does not depend on paid search content - there is no arrow between $P$ and $O$. This is true for some search engines like Google \citep{adwords2016tutorial}, but may not hold for other search engines.

%%In the next subsections we extend the above causal diagram from the query level to the aggregate level in various common scenarios w.r.t. search ads.

%%Figure~\ref{fig:A-causal-diagram} is an example with five nodes representing the following causal relationships: the value of node $Y$ is determined by $O$ and $P$, $O$ is determined by $Q$ while $P$ is determined by both $Q$ and $B$. Therefore intervention on $P$ has direct effect on $Y$, while intervention on $B$ does not have effect on $Y$ unless it causes changes in $P$.

In observational studies like MMM, measurements are often only possible for some of the nodes in the causal diagram. In order to measure the causal effect of ad spend on sales, it is important to first understand the underlying causal diagram, and then judge whether the causal effect is identifiable from the partially observed data. The back-door criterion originated by \citet{pearl1993bayesian} provides some theoretical guidance for this. To make the paper self-contained, we briefly review the relevant theory in the next subsection.

\subsection{Pearl's causal framework}

Pearl's description of causal diagrams as models of intervention are
important to understanding the concept of causal identifiability that
we use. Each child $X_i$ in a causal diagram represents a relationship
\begin{eqnarray}
  X_i &=& f_i(pa_i,\epsilon_i) \label{eq:pearl_eq}
  \end{eqnarray}
where $f_i$ is a function, $pa_i$ is the set of parents of $X_i$
and $\epsilon_i$ is an arbitrarily-determined random disturbance that
must be independent of all other variables and disturbances in the model.

\begin{defn*} (Causal effect, Pearl 2013) Given two variables, $X$ and $Y$,
the causal effect of $X$ on $Y$, denoted $\text{Pr}(y \mid \check{x})$, is a function
 from $X$ to the space of probability distributions on $Y$. For each
 realization $x$ of $X$, $Pr(y \mid \check{x})$ gives the probability of $Y=y$
 induced by deleting from model \eqref{eq:pearl_eq} the equation corresponding
 to $X$ and forcing $X$ to equal $x$ in the remaining equations. The $\check{x}$
 notation indicates ``intervene by setting $X$ to $x$''.
\end{defn*}

\begin{defn*} (Identifiability, Pearl 2013) The causal effect of $X$ on $Y$ is
identifiable if the quantity $\text{Pr}(y \mid \check{x})$ can be computed uniquely
from any positive probability of the observed variables that is compatible
with the diagram.
\end{defn*}

Identifiability means that, given an arbitrarily large sample from the joint distribution described by the causal diagram, the causal effect $\text{Pr}(y \mid \check{x})$ can be determined.

\begin{defn*} (d-separation, Pearl 2013) A path between two nodes on a causal diagram is said to be d-separated or blocked by a subset of variables (nodes) $Z$ if and only if either of the two conditions is satisfied: 1) the path contains a chain $i\rightarrow m\rightarrow j$ or a fork $i\leftarrow m\rightarrow j$ such that $m\in Z$, or 2) the path contains an inverted fork $i\rightarrow m\leftarrow j$ such that $m\notin Z$ and such that no descendant of $m$ belongs to $Z$.
\end{defn*} Now the back-door criterion can be stated as follows.

\begin{defn*} \label{def:back-door}(The back-door criterion, Pearl
2013) Given a causal diagram, a set of variables Z satisfies the back-door
criterion relative to an ordered pair of variables ($X$, $Y$) in
the diagram if: 1) no node in $Z$ is a descendant
of $X$; and 2) $Z$ ``blocks'' every path between $X$ and $Y$
that contains an arrow into $X$. \end{defn*}

Condition 1) in the definition of the back-door criterion rules out
covariates which are consequences of $X$, and condition 2) makes
sure that $Z$ contains the right set of confounding factors.
The back-door adjustment
theorem \citep{pearl2013causality} says that if a set of variables
$Z$ satisfies the back-door criterion
relative to $(X,Y)$, then the causal effect of $X$ on Y is
identifiable and the causal effect of $X$ on $Y$ is given by the formula
\begin{eqnarray}
  \text{Pr}(Y \mid \check{x}) = \sum_z \text{Pr}(Y \mid x, z)\text{Pr}(z).\label{eq:causal_estimate}
\end{eqnarray}
In other words, $Z$ makes it possible to estimate the causal effect of $X$ on $Y$.

In the example described by Figure \ref{fig:A-causal-diagram}, since there is only one path from $P$ to $Y$ that has an arrow into $P$, i.e. $P\leftarrow Q\rightarrow O\rightarrow Y$, obviously the node $Q$ (search query) meets the back-door criterion for the causal effect of node $P$ on $Y$. This makes it possible to estimate the causal impact of search ad given proper query level data; \citet{liu2012click} reported some pioneer work in this direction.

Pearl's framework has the same goal as and can be translated to
the counterfactual framework defined in the Neyman-Rubin causal model
\citep{holland1986}, but it also provides  formal semantics to help
visualize causal relationships.
See \citet{pearl2013causality} for detailed discussion.
The back-door criterion provides a convenient
tool for us to identify the proper set of covariates which satisfies
the so-called ignorability assumption in order to identify
causal effects from observational data \citep{rosenbaum1983}.
A general identification condition for causal effects has been developed
in \citep{tian2002general,maathuis2015generalized}. Our methodology of
selection bias correction for search ads is based on the back-door criterion and the assumption that ad serving has a random component.

Note that Pearl's framework puts aside three major questions that we have to
address in order to use it.  First, how to construct the causal diagram? Second,
can all necessary variables be measured accurately, even if they are observable?
Third, given finite sample size, what is the functional form of $\text{Pr}(Y \mid X,Z)$
when identifiability has been established as in Eq
\eqref{eq:causal_estimate}? The first question requires deep domain knowledge.
The second question may be addressed by careful data validation.
The last question may be alleviated when sample size is
sufficiently large to allow for non-parametric estimates, but in ads measurement, and especially in MMM, datasets are often quite small and so these practical considerations matter a lot.

\section{\label{sec:Methodology}Methodology}

With a focus on overall budget allocation across channels, the standard industry
MMM takes as a given a causal diagram where details of page ranking and the ad
auction are ignored. Since search ad spend and exposures are actually
intermediate outcomes influenced by bids, budget and consumer click behavior,
the standard MMM problem is inherently mis-specified for search. We take the
standard MMM problem as a given and show that reasonable results may be obtained
even with this misspecification. We briefly discuss a more realistic causal
diagram for search in the Appendix.

We formulate the ROAS problem by starting with simple cases where search ad is
the only media channel that an advertiser has invested. Under some realistic
assumptions, we use the back-door criterion to derive the method of bias
correction for the corresponding causal diagram. The theory and method is then
extended to more complex cases.

\subsection{\label{subsec:simple} Simple scenario}

In the simple scenario, search advertising is assumed to be the only advertising channel, and the contribution of other media channels on sales, if any, is ignorable. Let $X_{t}$ be the search ad spend for a particular product sold by an advertiser at time window $t$ and $Y_{t}$ be sales for the product during time window $t$. We assume
that the impact of search ads on sales occurs within the same period as the ad
exposure.

Consider the model below:
\begin{eqnarray}
Y_t & = & \beta_{0}+\beta_{1}X_t+\epsilon_t\label{eq:simple-roas}
\end{eqnarray}
where the parameter of interest is $\beta_{1}$, measuring the expected incremental
value of one unit change in search ad spend $X_t$ but conditional on no change in
$\epsilon_t$. Here $\beta_{1}$ is called the ROAS for search ads. That
is, $\beta_{1}X_t$ measures the causal impact of search ads on sales, and
$\epsilon_t$ represents other impact on sales (with the mean absorbed by the
intercept $\beta_{0}$) which are not explained by $X_t$.

The major factor which prevents us from obtaining unbiased estimates of $\beta_{1}$ by ordinary least squares (OLS), is the correlation between $X_{t}$ and
$\epsilon_{t}$. This is called the endogeneity problem in econometrics.
Throughout the paper, we drop the subscript $t$ if it causes no confusion.

In fact, by rewriting $\epsilon=\gamma X+\eta$, with
$\gamma=\text{cov}(X,\epsilon)/\text{var}(X)$ and $\eta=\epsilon-\gamma X$, we have
\begin{eqnarray*}
Y & = & \beta_{0}+(\beta_{1}+\gamma)X+\eta.
\end{eqnarray*}
It is easy to verify that $\text{cov}(X,\eta)=0$ and thus the naive estimate $\hat{\beta}_1$ through OLS has expectation $\beta_{1}+\gamma$ instead of $\beta_{1}$.

To obtain an unbiased estimate of $\beta_{1}$, it is critical to understand
what $\epsilon$ consists of. An important contributor to sales is the
direct impact from underlying consumer demand, denoted as $\epsilon_{0}$, which
can be affected by economic factors and seasonality. Organic search results may
contribute directly to sales, denoted as $\epsilon_{1}$. Due to ads targeting,
organic search content and paid search content are typically
positively correlated, resulting in $\text{cov}(X,\epsilon_{1})>0$. To be pragmatic, we
model the main effect as in \eqref{eq:simple-roas}.
It is often expected that $\text{cov}(X, \epsilon_0) >0$ and thus $\text{cov}(X, \epsilon) > 0$ if $\epsilon=\epsilon_{0}+\epsilon_{1}$,
which explains the phenomenon of over-estimation by the naive regression.

Let $V$ be the sufficient statistics to summarize the number of relevant
search queries that have potential impact on the sales of the product. Since different queries
may have a different effect on sales, $V$ is measured as a multi-dimensional
time series. Detailed implementation for deriving $V$ is left to
Section~\ref{sec:Implementation}. When $V$ is measured accurately, based
on the search ads mechanism described in Section \ref{subsec:causal.diagram} it is reasonable to assume that
\begin{eqnarray}
\epsilon_{1} & \perp & X \mid V\label{eq:paid_organic_CIA}
\end{eqnarray}
i.e. conditional on the relevant search queries, search ad spend is independent of potential
organic search impact.

Recall that search ads are determined by two parts: search queries are available to match keywords targeted by the advertiser; the advertiser has the budget to participate in the auction for search ads. To derive a working example causal diagram, we make two simple and explicit assumptions as follows:

(a) the advertiser's budget for search ads is unconstrained, and

(b) conditional on volumes of relevant search queries, the impact of consumer demand or
other economic factors on auction such as the advertiser's bid and competitors' actions is ignorable.

\begin{figure}
\begin{centering}
\includegraphics[width=4.5in,height=3in]{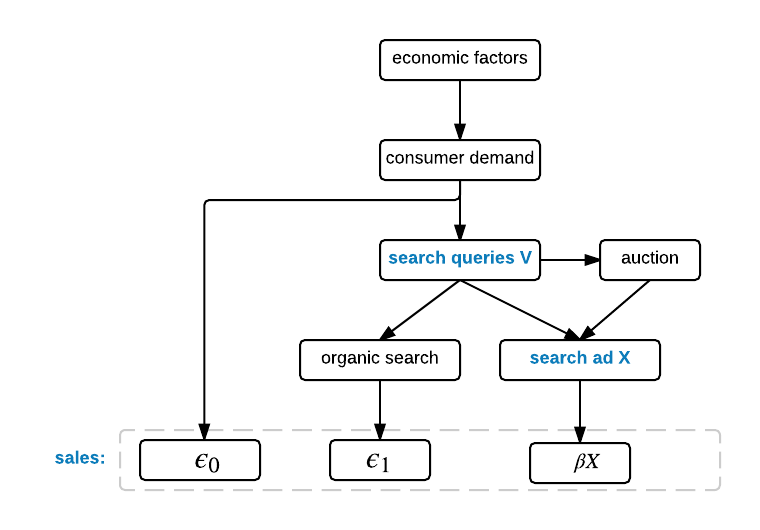}
\par\end{centering}

\caption{\label{fig:Causal-diagram-simple}Causal diagram for paid search (simple
scenario), where $X$ represents search ad spend; A more realistic causal diagram for search ad spend is given in the Appendix.}
\end{figure}

Under these assumptions, the causal diagram can be described as in Figure
\ref{fig:Causal-diagram-simple}.
The diagram implicitly assumes both \eqref{eq:paid_organic_CIA} and
\begin{eqnarray*}
\epsilon_{0} & \perp & X \mid V.
\end{eqnarray*}

The assumptions above are not unrealistic. Though an advertisers' budget is always finite, it is quite common\footnote{https://support.google.com/adwords/answer/2375418?hl=en} that advertisers rely on bid optimization instead of specific budget constraint to control search ad spend, under which assumption (a) holds.
Assumption (b) may be harder to verify but we suspect it holds in general if advertisers follow the bid strategy described by \citet{varian2009online}. Furthermore, the assumptions are just examples, under which it is relatively easier to verify or reject the causal diagram; the assumptions can be relaxed. We consider the scenario depicted
by Figure \ref{fig:Causal-diagram-simple} to be the simple scenario.

\begin{thm} \label{thm:simple-scenario}Assume that the causal diagram in Figure~\ref{fig:Causal-diagram-simple}
for paid search holds. If $X$ and $V$ are not perfectly correlated, then under regularity conditions\footnote{See \citet{bickel1998efficient} for the definition of regularity conditions for semiparametric models.}, search ad ROAS, i.e. $\beta_{1}$ in model \eqref{eq:simple-roas} can be estimated consistently by fitting the additive regression model below:
\begin{eqnarray}
Y & = & \beta_{0}+\beta_{1}X+f(V)+\eta\label{eq:theorem-simple-ssbc}
\end{eqnarray}
where $f(\cdot)$ is an unknown function and $\eta$ is the residual,
uncorrelated with $X$ and $f(V)$. \end{thm}

\begin{proof} There are four paths from search ad spend $X$ to sales that
  contains an arrow into search ad as shown in Figure
  \ref{fig:Causal-diagram-simple}:
  $X\leftarrow V \rightarrow \text{organic search} \rightarrow\epsilon_{1}$,
  $X\leftarrow \text{auction} \leftarrow V \rightarrow \text{organic search} \rightarrow\epsilon_{1}$,
  $X\leftarrow V\leftarrow\text{consumer demand}\rightarrow\epsilon_{0}$, and
  $X\leftarrow \text{auction} \leftarrow V\leftarrow\text{consumer demand}\rightarrow\epsilon_{0}$.
It is easy to check that $V$ satisfies the back-door criterion
relative to search ad and sales. According to the back-door adjustment
theorem, the causal effect of $X$ on $Y$ is identifiable by $(Y,X,V)$.

Let $f(v)=E(\epsilon \mid V=v)$ and $\eta=\epsilon-E(\epsilon \mid V)$. Now
according to model \eqref{eq:simple-roas}, the average causal effect
can be identified from conditional expectation:
\begin{eqnarray*}
E(Y \mid X,V) & = & \beta_{0}+\beta_{1}X+E(\epsilon \mid X,V).
\end{eqnarray*}
Due to the conditional independence $(\epsilon_{0},\epsilon_{1})\perp X \mid V$ assumed
by the causal diagram, we have
\begin{eqnarray*}
E(\epsilon \mid X,V) & = & E(\epsilon \mid V).
\end{eqnarray*}
Then
\begin{eqnarray*}
E(Y \mid X,V) & = & \beta_{0}+\beta_{1}X+f(V).
\end{eqnarray*}

By the identifiability theorem of additive index models \citep{yuan2011identifiability},
both $f(\cdot)$ and $\beta_{1}$ are identifiable. Therefore, under
regularity conditions, $\beta_{1}$ can be estimated consistently by the
usual regression method which minimizes $ \mid  \mid Y-\beta_{0}-\beta_{1}X-f(V) \mid  \mid ^{2}$
w.r.t. parameters $(\beta_{0},\beta_{1},f)$ with proper regularization on $f$.
When $f$ is known to
be a linear function, the estimate of $\beta_{1}$ is not only consistent
but unbiased. \end{proof}

The model \eqref{eq:theorem-simple-ssbc} falls into the class of semi-parametric
models \citep{bickel1998efficient}, where the parameter of interest is
$\beta_{1}$ and the nuisance
parameters include $f(\cdot)$ and the residual distribution of $\eta$,
assumed to have mean 0 and unknown finite variance. The estimation
procedure is described in detail later. We note that even when the
causal effect of search ads deviates from the simple linear form,
the formulation \eqref{eq:simple-roas} may still provide interesting
insight regarding the average causal effect. The result can be extended
naturally when the linear form $\beta_{1}X$ is relaxed to an unknown
function, which is described in Section \ref{sec:Implementation}.

\textbf{Remark 1.}
Assumptions (a) and (b) above are special cases where one expects
the causal diagram in Figure \ref{fig:Causal-diagram-simple} to hold.
Assumption (a) is relatively easy to check. The essential assumption required
by the causal diagram is that search ad spend only depends on the
volumes of relevant search queries and
other factors can be treated as noise unaffected by consumer demand.

\textbf{Remark 2.}
The assumptions in Theorem \ref{thm:simple-scenario} are sufficient
but not necessary; for example, if search ad spend only depends on ad budget
and is entirely randomized so that assumption (a) is violated, then
model \eqref{eq:theorem-simple-ssbc}
can still give a consistent estimate of search ad ROAS as defined in
\eqref{eq:simple-roas}.

\textbf{Remark 3.}
There exists scenarios where the causal diagram in Figure
\eqref{fig:Causal-diagram-simple} does not hold. For example,
  weather has dramatic impact on both consumer demand and supply on the fish
  market \citep{angrist2000}. If weather becomes too bad, it may reduce both
  consumer demand and supply dramatically, then there can be a path
  from consumer demand to $X$ which does not go through search queries, but
  through weather and supply assuming that the supply market advertises on
  search through auction. In this scenario, search ad ROAS is not identifiable unless weather
  or supply is taken into account. See Section \ref{subsec:counter-examples}
  for a few more counter examples.

\subsection{\label{subsec:complex} Complex scenario}

Now we consider cases where search advertising is not the only
channel that may affect sales significantly. We let $X_{2}$ denote
all non-search ad contributors, e.g. traditional media channels and non-search
digital channels, which may directly affect sales. Non-search contributors
may also trigger consumers to search more online for the product
(i.e. a funnel effect). Advertisers might want to plan budgets for both
search ads and other media channels. We use the graph in
Figure \ref{fig:Causal-diagram-for-complex-1}
as an example of causal diagram for such a scenario. As in the case above, this graph
is a dramatic simplification. For example, it does not describe complexity such as historical ads may impact current sales (lag effect of non-search contributors), and it may ignore potentially weak links not shown on the diagram.
%%account for  lagged ad effects, among many other complexities. .

If search ad spend is not directly correlated with other media spend, but is mostly determined by the availability of search ad inventory through consumers' relevant search query volume, then
the causal diagram reduces to Figure \ref{fig:Causal-diagram-for-complex-2}.  This holds approximately for many advertisers, for example when advertisers use bid optimization instead of specific budget constraint to control search ad spend. Under this approximation, non-search contributors as well as their potential lag effects do not affect the identifiability of $\beta_1$.

We derive the simplified theory for the complex scenarios as in Theorem \ref{thm:complex-scenario}.

\begin{center}
\begin{figure}
\begin{centering}
\includegraphics[width=4.5in,height=3in]{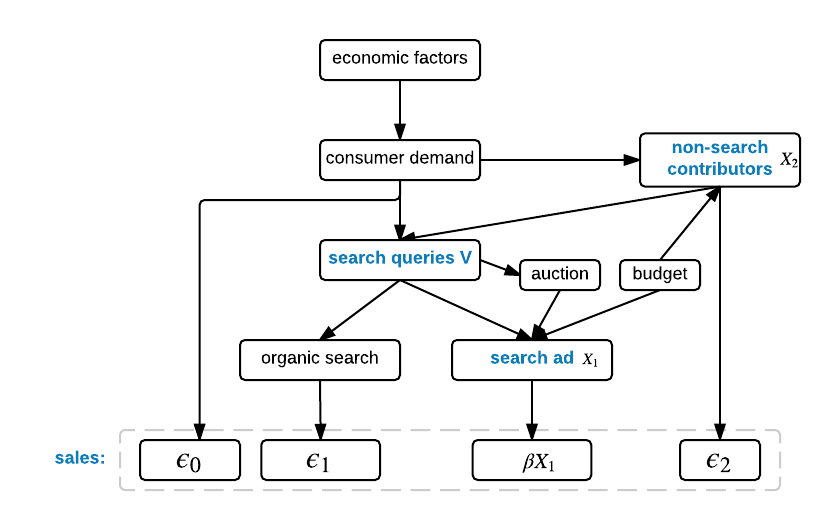}
\par\end{centering}

\caption{Cau\label{fig:Causal-diagram-for-complex-1}sal diagram for search
ad (complex scenario 1)}
\end{figure}

\par\end{center}

\begin{center}
\begin{figure}
\begin{centering}
\includegraphics[width=4.5in,height=3in]{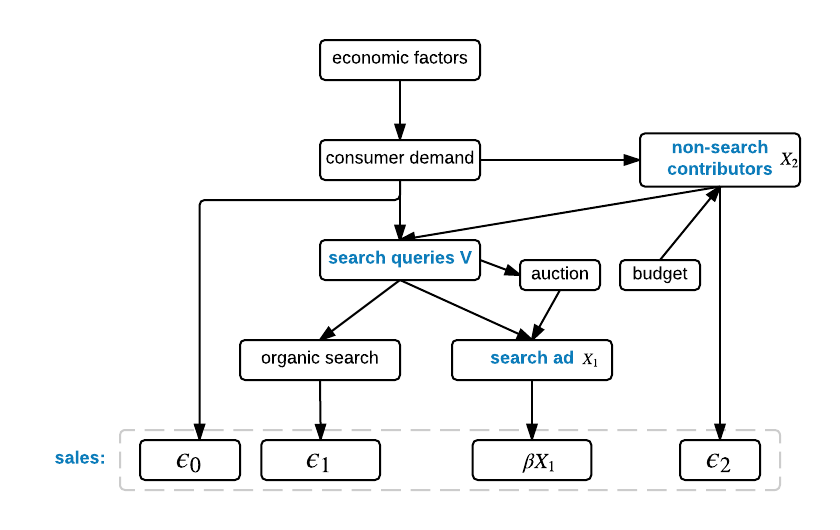}
\par\end{centering}

\caption{Causal diagram for search ad (complex scenario 2), where the only difference from Figure \ref{fig:Causal-diagram-for-complex-1} is the lack of arrow
from budget to $X_1$ due to unconstrained budget for search ad spend.}
\label{fig:Causal-diagram-for-complex-2}
\end{figure}

\par\end{center}

\begin{thm} \label{thm:complex-scenario}(1) Assume that the causal
diagram in Figure \ref{fig:Causal-diagram-for-complex-1} for search
ads holds and that $X_2$ has ignorable lag effect.
The causal effect of paid search on sales is identifiable
from observational data $(X_{1},X_{2},V,Y)$. If $X_{1}$ is not perfectly
correlated with $V$ and $X_{2}$, then under regularity conditions,
search ads ROAS $\beta_{1}$
defined in model \eqref{eq:simple-roas} can be estimated consistently
by fitting the additive regression model below:
\begin{eqnarray}
Y & = & \beta_{0}+\beta_{1}X_{1}+f(V,X_{2})+\eta\label{eq:theorem-complex-ssbc-1}
\end{eqnarray}
where
\begin{eqnarray*}
f(v,x_{2}) & = & E(\epsilon_{0} \mid V=v,X_{2}=x_{2})+E(\epsilon_{1} \mid V=v)+E(\epsilon_{2} \mid X_{2}=x_2)
\end{eqnarray*}
and $\eta$ is the residual, uncorrelated with $X_{1}$ and $f(V,X_{2})$.

(2) If the causal diagram in Figure \ref{fig:Causal-diagram-for-complex-2}
holds, then under regularity conditions,
search ad ROAS $\beta_{1}$
defined in model \eqref{eq:simple-roas} can be estimated consistently
by fitting the additive regression model below:

\begin{eqnarray}
Y & = & \beta_{0}+\beta_{1}X_{1}+f(V)+\eta,\label{eq:theorem-complex-ssbc-2}
\end{eqnarray}
where $\beta_{1}$ is the parameter of interest and $f$ is an unknown
function. That is, the estimation procedure is the same as for the
simple scenario described earlier.\end{thm}

\begin{proof} First prove (1). It is not hard to verify by definition that
$(V,X_{2})$ satisfies the back-door criterion for $X_{1}\rightarrow Y$
and thus makes the causal effect of $X_{1}$ on $Y$ identifiable.
Next due to $\epsilon_1 \perp X_1 \mid V$, $\epsilon_2 \perp X_1 \mid X_2$ and
$\epsilon_0 \perp X_1 \mid (V, X_2)$ assumed by the causal diagram,
one can show that
\begin{eqnarray*}
E(Y \mid X_{1},X_{2},V) & = & \beta_{0}+\beta_{1}X_{1}+E(\epsilon_{1} \mid V)+E(\epsilon_{2} \mid X_{2})+E(\epsilon_{0} \mid V,X_{2}).
\end{eqnarray*}

Result (2) can be proved similarly. \end{proof}

\textbf{Remark 4.}
As mentioned earlier, it is quite common that advertisers put no budget
constraint on search ad spend. This implies that the scenario identified
by Figure \ref{fig:Causal-diagram-for-complex-2} can be more common than
the more complex one identified by Figure \ref{fig:Causal-diagram-for-complex-1}.
Practical models for the scenario of Figure \ref{fig:Causal-diagram-for-complex-1} may require
careful consideration of lag effects in $X_2$.

\textbf{Remark 5.}
Note that $(X_{1},V)$ does not satisfy the back-door criterion for
$X_{2}\rightarrow Y$, since the path
$X_{2}\leftarrow\text{consumer demand}\rightarrow\epsilon_{0}$
is not blocked. For example, $X_2$ may represent social media ad spend.
This suggests that the causal effect of $X_{2}$ on
sales cannot be estimated consistently by observations on $(Y, X_1, X_2, V)$
only.

\textbf{Remark 6.}
It may be worth pointing out that even if one may be able to collect
additional variables so as to satisfy the back-door criterion for
$X_{2}\rightarrow Y$, there is no guarantee that one can estimate
the causal effects of $X_{1}$ and $X_{2}$ simultaneously from a
single regression in traditional MMMs as described in \citet{jin2017}.
If the two subsets of variables that satisfy the back-door criterion for $X_{1}\rightarrow Y$ and
$X_{2}\rightarrow Y$ separately, are not the same, by regression against
all relevant variables one may obtain uninterpretable results and even
Simpson paradox. For example, by conditioning on unnecessary covariates, one may obtain negative impact for some media while the true impact is positive.

\subsection{\label{subsec:fullmmm} Estimation of full MMM}
Much of the focus of this paper thus far has been around the estimation of the impact of search ad ($X_{1}$). For a practitioner of MMMs, it is also required to estimate the impact of the non-search ad media ($X_{2}$). The remarks above note that it would be difficult to obtain general conditions under which it is possible to estimate $X_{2}$ consistently, especially if the modeler was to use a single regression model.  Even if the modeler was to use a fully graphical model, estimation of $X_{2}$ consistently would remain a challenge due to the conditions that need to be satisfied.

If the requirement still is to estimate the impact of both $X_{1}$ and $X_{2}$ in the MMM, then one possible approach would be to estimate the impact of $X_{1}$ first, with the bias correction method applied.  The impact of $X_{1}$ can then be fixed in the full MMM, and the impact of $X_{2}$ can be fitted via traditional means such as described in \citet{jin2017}.  The modeler should view the estimated parameters for $X_{2}$ fitted via this approach with the same critical lens as if the bias correction method was not applied at all.

\section{\label{sec:Implementation}Implementation}

In this section, we first describe how to collect search query data $V$, which is not available in standard MMM data collection, and then describe the model fitting procedure.

\subsection{Summarization of search query data}\label{subsec:query-volume}

As noted in the previous section, $V$ represents the volumes of relevant search queries that have potential impact on the sales of the product. The total number of relevant search queries is potentially very large, so it is important to summarize search queries in a way that can be used conveniently for model fitting. The summarization of $V$ is not straightforward, as the potential impact of each query term can be different. Below we describe a procedure to summarize search queries based on their potential impact on organic search results.
%%However it may be possible to group together query terms such that their relevance to the product over time is similar. The procedure outlined below represents one possible way of summarizing the search query volume in a way that is tractable for the regression.  There are potentially other summarization approaches that could be applied, but we have found the approach below works well in practise.

{\bf Step 1} \\
Identify the advertiser's website and its top competitors's websites.

{\bf Step 2} \\
Collect all queries over a target region (e.g. US) in a given time window (e.g. last six months). For each query, count the number of times each URL appears in the organic search results.  These URLs are called destination URLs.  The data structure looks like this:

($\text{q}_i$, $\text{u}_{j}$, $n_{i,j}$)

($\text{q}_i$, $\text{u}_{j+1}$, $n_{i,j+1}$)

...

where $n_{i,j}$ is the number of times the $j$th URL appears with the $i$th query term.  Given query $\text{q}_{i}$, if the set of URLs associated with the query contains the advertiser's website, then the query is considered relevant to that advertiser.  Let $\mathcal{S}$ be the set of relevant queries.  Each relevant query in $\mathcal{S}$ may represent a  different level of demand for the advertiser's product.

{\bf Step 3} \\
Partition the relevant query set $\mathcal{S}$ into three groups according to the mix of URLs that appear for each query.  The destination URLs appearing in the organic results can be classified into four groups: a) belongs to the advertiser, b) belongs to top competitors, c) does not belong to the advertiser or its competitors, but belongs to the business category, and d) does not belong to the business category.

For any query $\text{q}_{i}$, the sum of the number of impressions for the URLs classified into each group can be denoted as $w_{i,a}, w_{i,b}, w_{i,c}$ and $w_{i,d}$ respectively.  Let $w_{i,\text{total}} = w_{i,a} + w_{i,b} + w_{i,c} + w_{i,d}$ be the total impressions for $\text{q}_{i}$ and $w_{i,\text{category}} = w_{i,a} + w_{i,b} + w_{i,c}$ be the category impressions for $\text{q}_{i}$.

If $w_{\text{category}}/w_{\text{total}}$ is less than a pre-determined threshold, ignore the query as it is less likely to be relevant to the business category.

Otherwise: if $w_{a}/w_{\text{category}}$ is greater than a pre-determined threshold, classify it as target-favoring, else if $w_{b}/w_{\text{category}}$ is greater than a threshold, classify it as competitor-favoring, else classify it as general-interest.

This gives us three subsets of queries, say, $\mathcal{S}_{1}$ containing all target-favoring queries, $\mathcal{S}_{2}$ containing all competitors-favoring queries, and $\mathcal{S}_{3}$ containing all general-interest queries.

{\bf Step 4} \\
Given the three sets of queries $\mathcal{S}_{1}, \mathcal{S}_{2}$ and $\mathcal{S}_{3}$, we can count the total number of searches for each query set in each time window $t$ and label it  as $V_{1t}$ (target-favoring), $V_{2t}$ (competitors-favoring) and $V_{3t}$ (general-interest) correspondingly. The sum $V_{1t} + V_{2t} + V_{3t}$ is called category search volume at time window $t$.

\begin{figure}
\begin{centering}
\includegraphics[width=2.5in,height=2.5in]{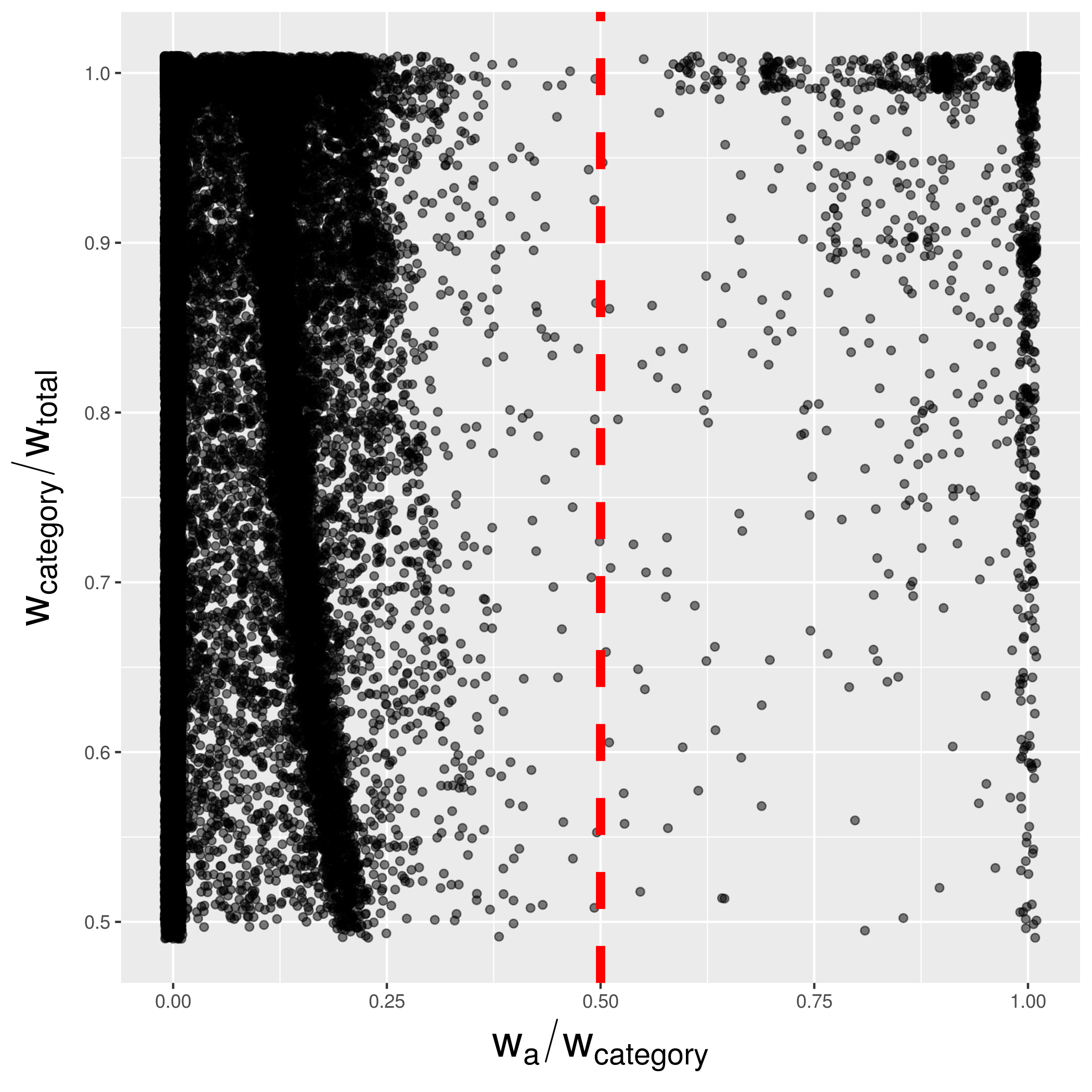}\includegraphics[width=2.5in,height=2.5in]{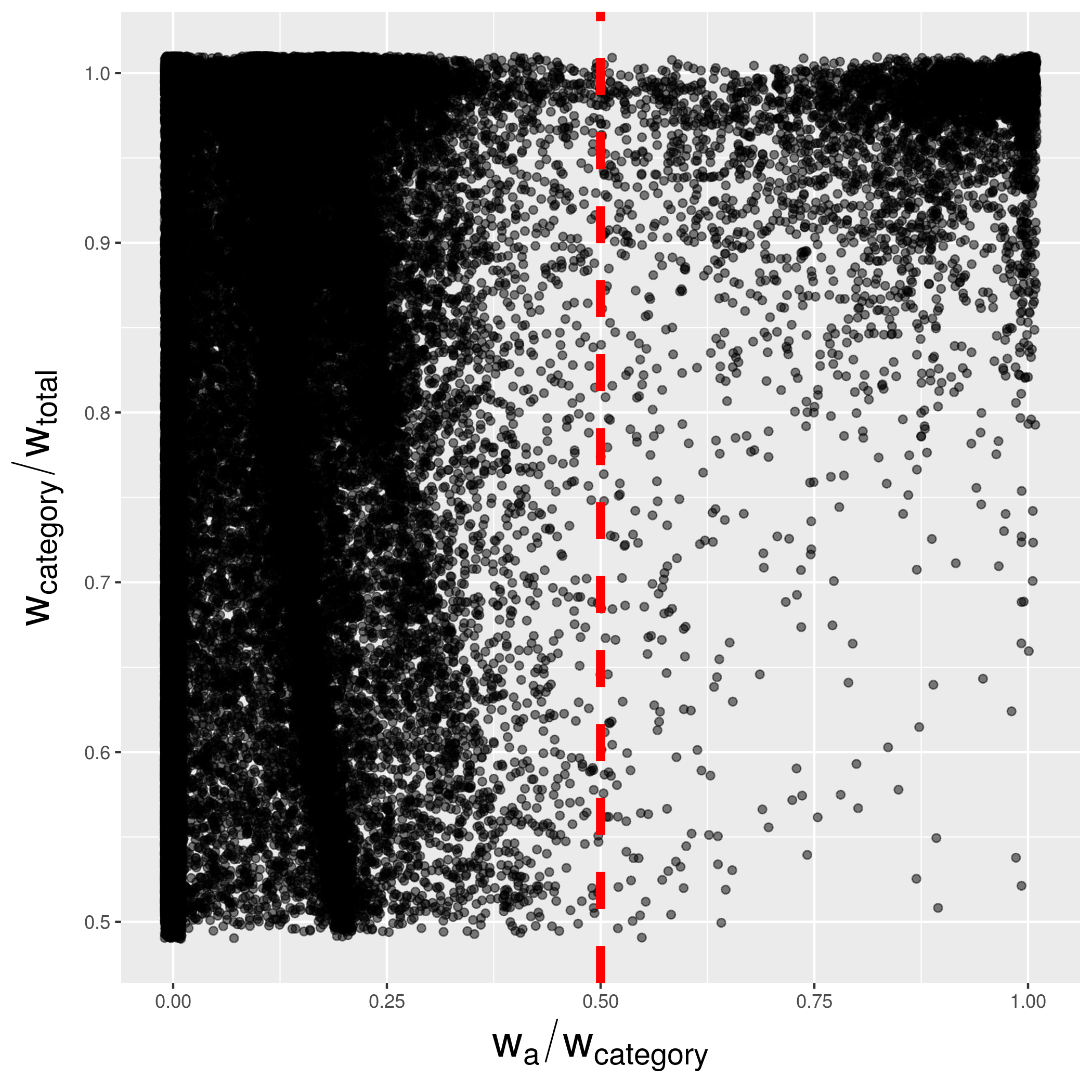}
\par\end{centering}

\begin{centering}
\includegraphics[width=2.5in,height=2.5in]{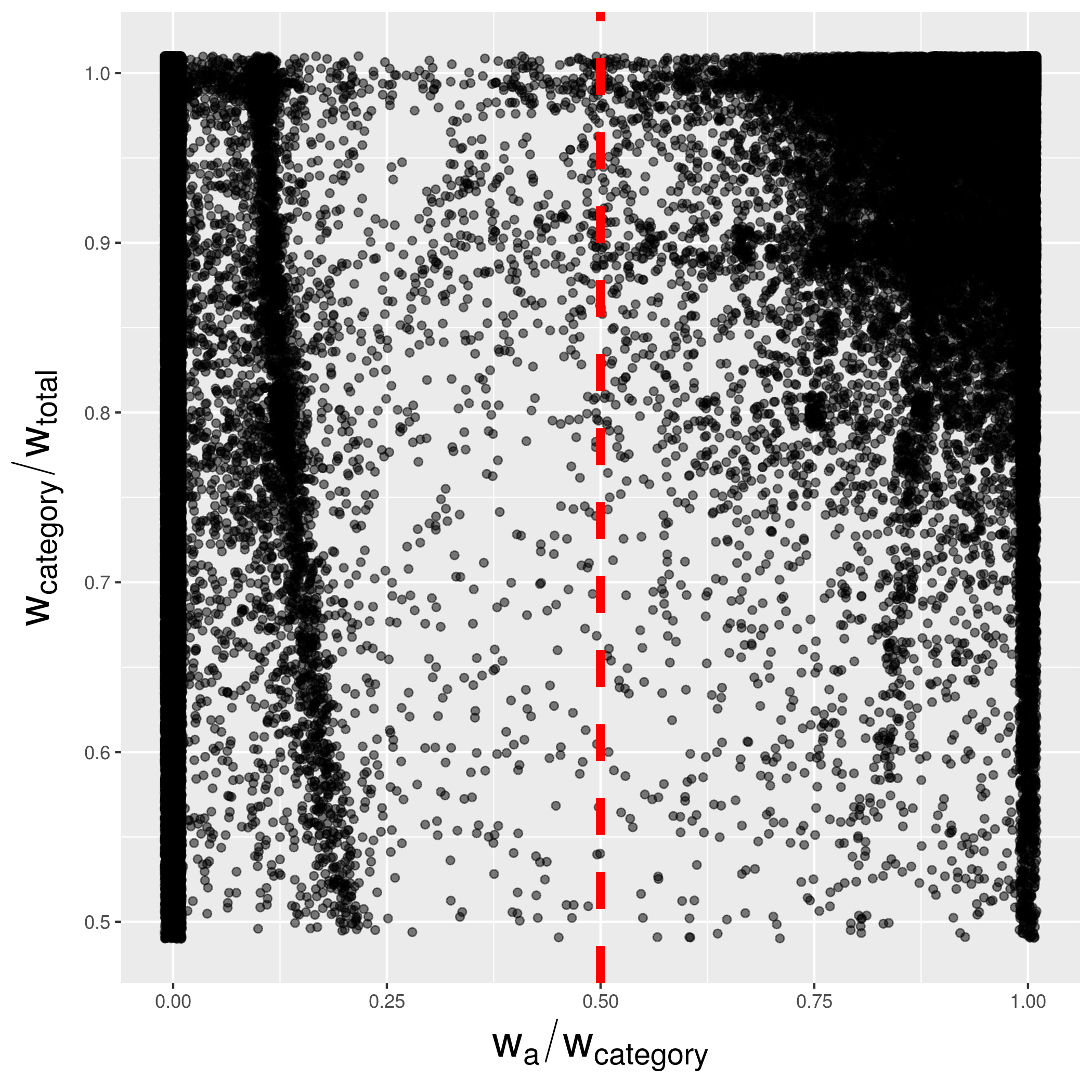}\includegraphics[width=2.5in,height=2.5in]{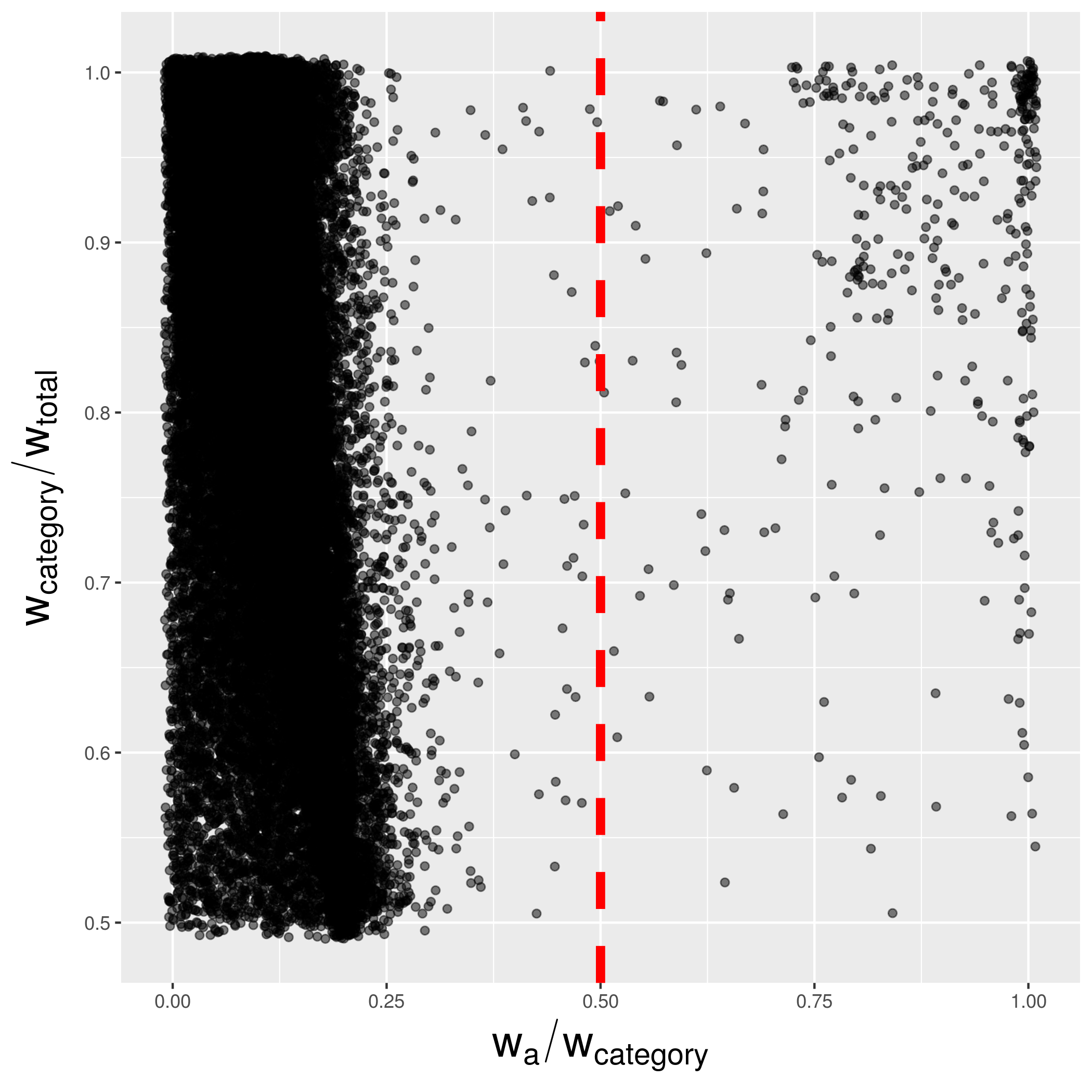}
\par\end{centering}

\caption{\label{fig:query-classification}
  Examples of search query classification, where each dot is for a relevant
  query, x-axis shows the ratio $w_a/w_{\text{category}}$ and y-axis shows the ratio
  $w_{\text{category}}/w_{\text{total}}$ (see Step 3 in Section \ref{subsec:query-volume}
  for the definition of $w_a$, $w_{\text{category}}$ and $w_{\text{total}}$);
  In each case, queries on the right hand side of each
  vertical red line are grouped as target-favoring.}
\end{figure}

Empirically we have found that 50\% is a reasonable choice
for the thresholds required for the above segmentation procedure.
Figure~\ref{fig:query-classification} shows the scatter plots of queries
in terms of $w_{a}/w_{\text{category}}$ and $w_{\text{category}}/w_{\text{total}}$ for
four different case studies, which show clusters on both sides of the vertical line at 50\%.
Note that the segmentation procedure is based on domain-knowledge of the advertiser and
the related queries, and could probably be refined.

\subsection{Model fitting procedure}

Implementation of our SBC method relies on fitting the additive models identified by Theorem \ref{thm:simple-scenario} and Theorem \ref{thm:complex-scenario} in Section \ref{sec:Methodology}.

For the simple scenario, we approximate the function $f(V)$ defined in Theorem \ref{thm:simple-scenario} by an additive function $\sum_{i=1}^{3}f_{i}(V_{i})$, where $V=(V_{1},V_{2},V_{3})$. The bias corrected estimation of $\beta_{1}$ can be implemented by fitting an additive regression model \citep{hastie1990generalized} through the R function {\em GAM} in the library MGCV \citep{wood2012mgcv} as below:
\begin{eqnarray}
Y & \sim & \beta_{0}+\beta_{1}X+s(V_{1})+s(V_{2})+s(V_{3})\label{eq:fit-ssbc-simple}
\end{eqnarray}
where $s(\cdot)$ is the smooth function as described in \citet{wood2006low}.

We adopt the REML algorithm proposed by \citet{wood2011fast} which reformulates the additive regression procedure as fitting a parametric mixed effect model, and is already implemented in the library MGCV  \citep{wood2012mgcv}.
%%We call this procedure SBC--SP, where SP is short for semi-parametric.
Both point estimate and standard error are reported by {\em GAM}.

When the number of observations is large enough (which in this paper applies specifically to the case study in Section \ref{sec:case4}), instead of approximating $f(V)$ by an additive function, one can approximate $f(V)$ directly by a 3-dimension full tensor product smooth as described in \citet{wood2006low} and estimate $\beta_1$ by the regression below:
\begin{eqnarray}
Y & \sim & \beta_{0}+\beta_{1}X+te(V_{1}, V_2, V_3)\label{eq:fit-ssbc-simple-te}
\end{eqnarray}
where $te$ is the R function in MGCV to implement the full tensor product smooth.

To check model stability, we have also looked at results which replace $\beta_{1}X$ by an unknown smooth function $s(X)$, assumed to be monotonically increasing. The results were calculated based on marginal ROAS as defined in \citet{jin2017}, i.e.
\begin{align*}
	\hat{\beta}_{1} & =\sum_{t}(\hat{s}((1+\delta)X_t)-\hat{s}(X_t))/(\delta\sum_{t}X_t)
\end{align*}
This is a non-parametric model fitting procedure. In all case studies below, marginal ROAS point estimates by this procedure are very much comparable to the estimates from \eqref{eq:fit-ssbc-simple} but we have not evaluated the standard errors. Details may be reported in future work.

For the purpose of comparison, we also report the naive estimate fitted by OLS as follows:
\begin{eqnarray}
Y & \sim & \beta_{0}+\beta_{1}X.\label{eq:fit-biased-simple}
\end{eqnarray}

Consumer demand has a large impact on sales but it is hard to measure directly.  Modelers sometimes use proxy variables to control for the underlying consumer demand, so we also include the demand-adjusted estimate below for comparison, also fitted by {\em GAM}:
\begin{eqnarray}
	Y & \sim & \beta_{0}+\beta_{1}X+s(S)\label{eq:fit-seasonality-simple}
\end{eqnarray}
where $S$ stands for a consumer demand proxy variable. In the case studies below, category search volume is used for $S$.

For the complex scenario described by the causal diagram in Figure \ref{fig:Causal-diagram-for-complex-2} where there is no direct correlation between search ad spend and other media spend, it reduces to the simple scenario according to Theorem
\ref{thm:complex-scenario}. %% Thus both SBC--SP and SBC--NP can be applied here.

For the causal diagram in Figure \ref{fig:Causal-diagram-for-complex-1}, where there is correlation between search ad spend and other media spend induced by budget constraints and unblocked by any observable variable, the method described in \eqref{eq:theorem-complex-ssbc-1}
may be insufficient as we may need to consider lag effects, especially
for traditional media such as TV and direct mail. How to model long-term
lag effect is still an active open problem in the literature \citep{wolfe2016}.
Further research is required for the scenario identified by Figure \ref{fig:Causal-diagram-for-complex-1}.

\section{Case studies in simple scenarios}\label{sec:case123}

To understand the performance of the proposed SBC method in measuring search ad effectiveness, it is important to study real cases and compare with ground truth. It is not easy to collect the right data in practice. Fortunately we have been able to identify various cases where we have access to both media spend data and outcome metrics. These cases span from simple scenarios where search ads are known to be the dominating media channel, to a complex scenario, with more than a dozen media channels, including search ads.

In this section, we report three case studies from three different verticals which all fall into the simple scenario where search ads are the dominant media channel in terms of spend, and other media spends are much smaller.\footnote{We were able to identify four such cases in total, but the fourth case showed strong lag effect in search ad, requires a more complex model, and thus is not reported in this paper. More case studies may be reported in the future.} In each case, the advertiser ran a randomized geo-experiment to estimate the effect of their search ads.
%%Randomized geo experiments were carried out independently by Google Analytics.
\footnote{There are about 200 DMAs in the United States, defined by the Nielsen company. DMAs
are first paired according to comparable demographics and then DMAs
in each pair are randomly assigned to the control group or the treatment
group. See \citet{kerman2017tbr} for the estimation of search ad ROAS from randomized geo experiments.}

We use experimental results as the source of truth to compare to observational results. For each of the case studies, we compare various estimation methods: the naive estimate (NE), demand adjustment by category search volume (SA), the SBC method as described in Section \ref{sec:Implementation}.
In each of the three case studies, the data include overall search ad spend, the KPI and search query volumes in the U.S. on the daily basis over a few months. Both search ads spend and KPIs were reported by the clients, while search query volumes were collected
internally as described in Section \ref{subsec:query-volume}. The outcome variable (KPI) varies across experiments.  In case 1, the KPI was offline transaction value; in case 2 it was the number of inquiries; and in case 3 it was number of site visits. The ROAS values we report are on the scale of KPI/search dollar.
\begin{center}
\begin{figure}
\begin{centering}
\includegraphics[scale=0.5]{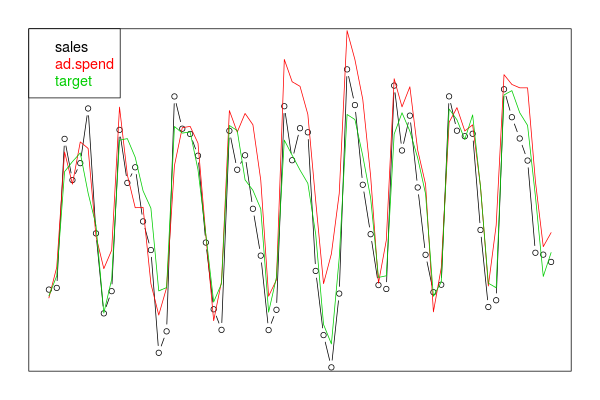}
\par\end{centering}

\caption{\label{fig:TS_simulated}Time series of sales, search ad.spend and
search query volume (the target-favoring dimension), simulated from real data
in the first case study below, where each time series is rescaled
by its median value.}
\end{figure}

\par\end{center}

The time series of each variable in each case follows a clear seasonality
pattern, e.g. day of the week, and seasonal trends -- see Figure
\ref{fig:TS_simulated} for an example which were simulated from one of the
cases. To keep data privacy, we do not report the scale of each variable, but
report some high level summary statistics such as pairwise correlation and
fitted model parameters.  Also, for each case study, the experimental point
estimate is scaled to equal one and all results and standard errors are
indexed to that result.

\begin{figure}
\centering
\begin{minipage}{0.75\textwidth}
\subfigure[Case 1]{\includegraphics[width=2in,height=2in]{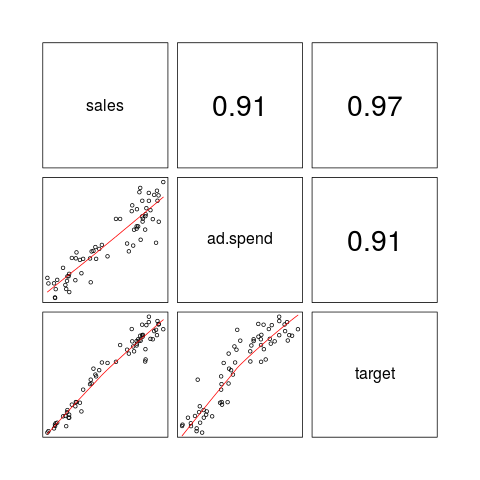}\includegraphics[width=2.5in,height=1.8in]{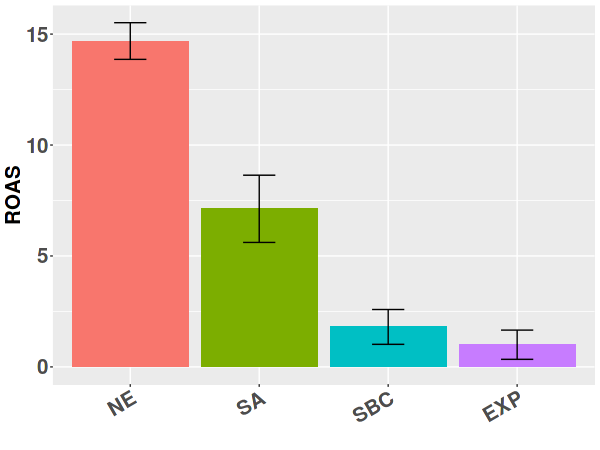}}

\subfigure[Case 2]{\includegraphics[width=2in,height=2in]{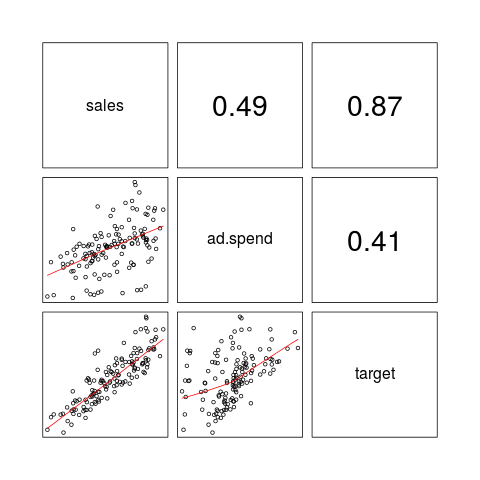}\includegraphics[width=2.5in,height=1.8in]{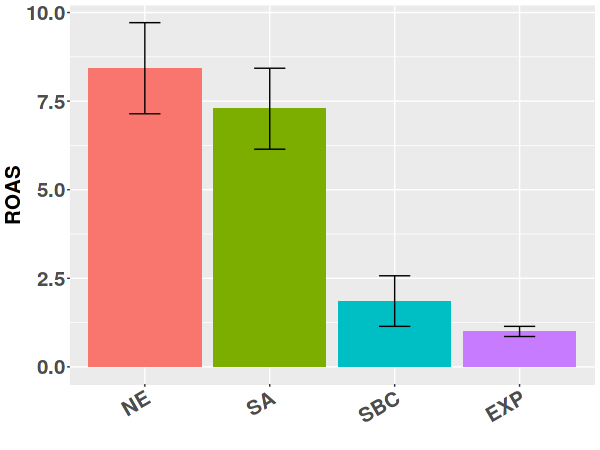}}

\subfigure[Case 3]{\includegraphics[width=2in,height=2in]{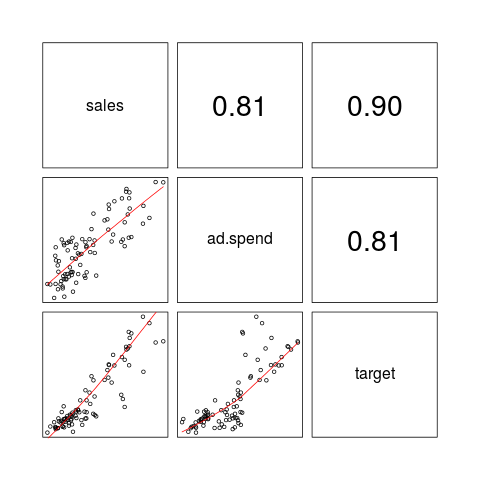}\includegraphics[width=2.5in,height=1.8in]{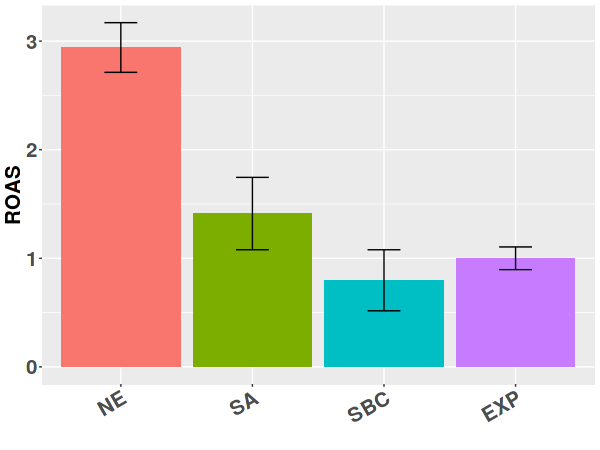}}

\end{minipage}
\caption{\label{fig:collinearity-simple}
  Report the pairwise scatter plots and correlations between search ad spend,
  target-favoring search volume and sales (Left panels) and estimated ROAS
  (Right panels) for the three case studies, where NE stands for the naive
  estimate and SA stands for the demand-adjusted estimate;
  EXP stands for the reference value
  from randomized geo experiments. The bar-lines show the values of
  $\hat{\beta}_{1}\pm std.error(\hat{\beta}_{1})$. Both point estimates and
  standard errors are rescaled by the original EXP point estimate
  in order to preserve data privacy.}
\end{figure}

\subsection{Case 1}

In this case, the advertiser is a medium-size
(with annual revenue of tens of millions of USD) %% e-commerce
retailer.
%%and provides offline delivery of the product after receiving customers' ordersonline.
Search advertising was the only major marketing channel, with
no significant spend on other media channels. We have daily metrics of sales,
ad spend and search query volumes for  65 days in 2015.
The left panel in the top row of
Figure \ref{fig:collinearity-simple} shows the pairwise scatterplot, where
the numbers on the upper panels are the Pearson correlation. For example,
the correlation between ad.spend and sales is 0.91. A simple linear
model with ad.spend can fit and predict sales well. The strong correlation
(0.91) between target-favoring search query volume and ad spend in this case
suggests that: 1) there
may be strong ad targeting, and 2) the advertiser rarely or never hits
the top of their search ad budget.
On the other hand, the correlation between search volume and sales
is 0.97.

\begin{figure}
\begin{centering}
\subfigure[Case 1]{\includegraphics[width=5in,height=2in]{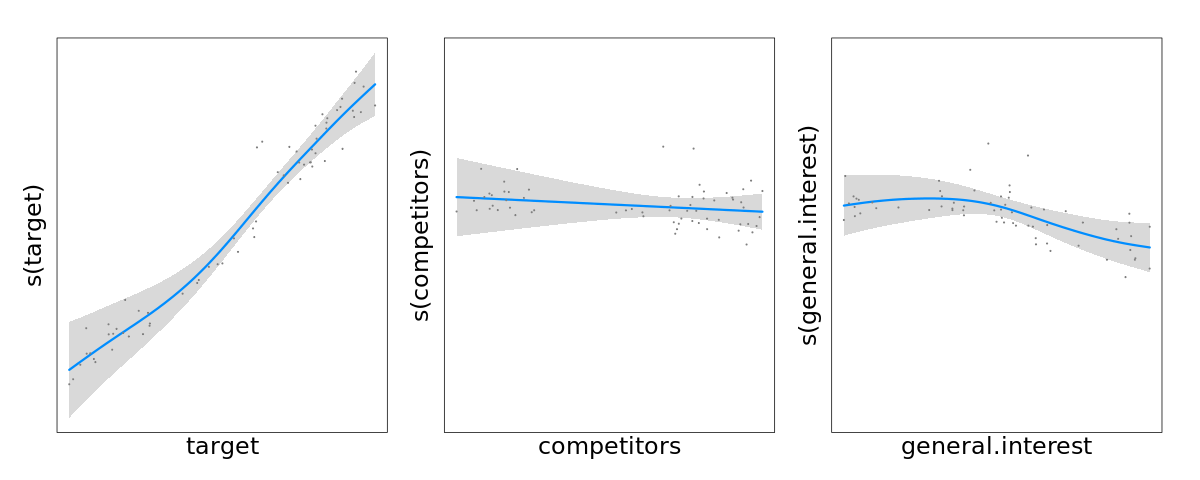}}
\par\end{centering}

\begin{centering}
  \subfigure[Case 2]{\includegraphics[width=5in,height=2in]{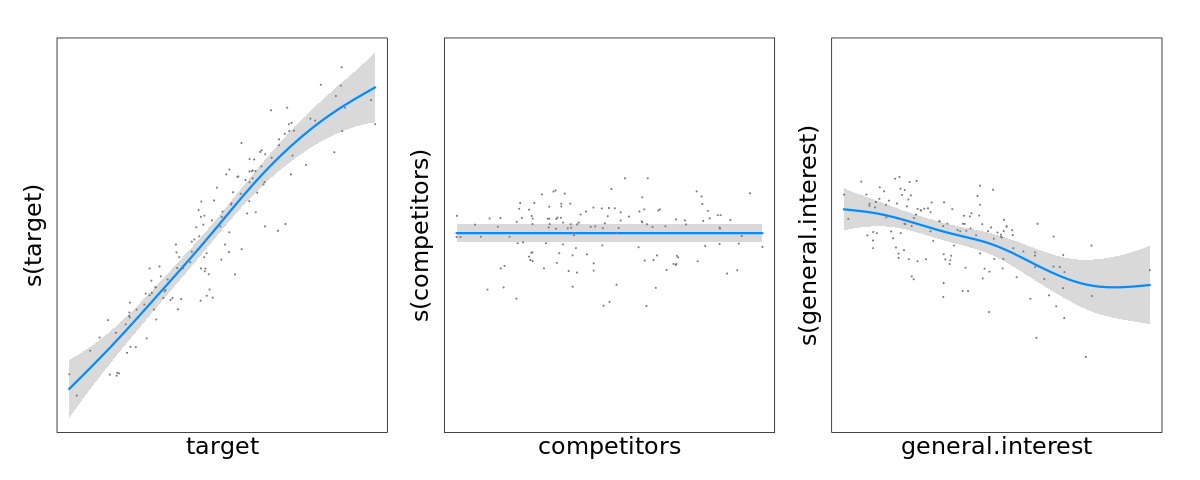}}
\par\end{centering}

\begin{centering}
  \subfigure[Case 3]{\includegraphics[width=5in,height=2in]{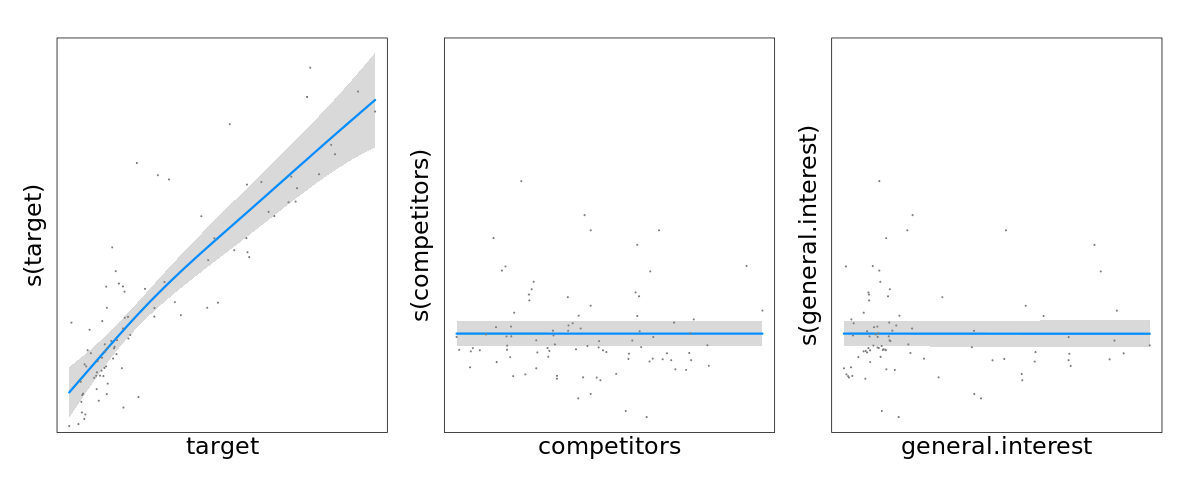}}
\par\end{centering}

\caption{Selection bias explained by changes of target-favoring, competitors-favoring and general interest search query volumes in Case 1, 2 and 3, where
  the response curves and 95\% confidence bands for the 3-dim search query
  volumes are fitted in an
  additive function as described in the regression \eqref{eq:fit-ssbc-simple};
  the scatter plots are fitted function values plus model residuals.}
\label{fig:Selection-bias-explained-case-123}
\end{figure}

First we fit SBC as described
in \eqref{eq:fit-ssbc-simple}:
\begin{eqnarray*}
\text{response} & \sim & \beta_0 + \beta_1 \times \text{ad.spend}+s(\text{target})+s(\text{competitors})+s(\text{general.interest})
\end{eqnarray*}
where target, competitors and general.interest represent target-favoring, competitor-favoring and general interest search query volumes separately.
The point estimate of $\beta_{1}$ is 3.0 with standard error 1.02. The fitted smooth function
for target-favoring query volume is monotonically increasing and
almost linear (see Figure \ref{fig:Selection-bias-explained-case-123}(a)).
The adjusted $R^{2}$ value is 0.95. The monotonicity is expected,
but it is interesting to see the fitted curve from data directly without
forcing monotonicity in any way. The fitted function for competitors-favoring
search volume on the other hand is pretty flat and is not statistically
significant, while the one for general interest is statistically significant.

The naive estimate of $\beta_{1}$ based on OLS (\ref{eq:fit-biased-simple})
is 14.7, with std.error 0.83.
Using category search volume to control for seasonal demand, as in model
(\ref{eq:fit-seasonality-simple}),
the fitted value is 7.1 with std.error 1.51. These two model fittings
have adjusted $R^{2}$ values of 0.83 and 0.90 respectively.

The advertiser conducted the randomized geo experiments during the second
month of the period.
The indexed experimental estimate of ROAS has std.error 0.66 . The naive
estimate of ROAS is almost 15-fold larger than the experimental result.
With the simple category-search-volume based demand adjustment, the gap
shrinks but the estimate is still seven times as large. In contrast,
the SBC estimate is much closer to the experimental result. See the comparison
in Figure \ref{fig:collinearity-simple}(a).

\subsection{Case 2}

In this case,
%%the advertiser's business is a travel marketplace.
the search ad spend, KPI,
search query volumes data are on a daily basis over a period of about 4 months
(135 days). The randomized experiment was carried out in the last 6 weeks.

In this case, the demand adjustment does not reduce the bias
much, bringing the estimated ROAS from 8.4 (with standard error 1.30) to 7.3 (with standard error 1.14).
On the other hand, the SBC estimate is 1.9 with standard error 0.71, much closer
to the experimental result with standard error 0.14.
See the comparison in Figure \ref{fig:collinearity-simple}(b).
The fitted smooth function for the target-favoring
search volume again is monotonically increasing and almost linear.
Like Case 1, the competitors-favoring search volume
is not statistically significant, as shown in
Figure \ref{fig:Selection-bias-explained-case-123}(b).
It is noticeable that the correlation between target-favoring search
volume and search ad spend is only 0.47, much lower than that
in Case 1, but the strong correlation between sales and search volume
may suggest that underlying consumer demand or organic search or both have
contributed to sales dramatically
in this case.

\subsection{Case 3}

In this case, %%the advertiser is a national apparel retailer and the
%%KPI is the  number of visits to their website.
the data covers about 3 months
(88 days) and the randomized experiment was carried out
in the last 6 weeks.

The SBC estimate of ROAS is 0.8 with standard error 0.28, the naive estimate is 2.9 with standard error 0.23, while the demand-adjusted estimate is 1.4 with standard error 0.33.
See the graphical comparison in Figure \ref{fig:collinearity-simple}(c).
In this case, the naive estimate is about three times larger than the
experimental result. The demand-adjusted estimate is about half of that,
much closer to the experimental result. As in Cases 1 and 2, taking into
account standard errors, the SBC estimate is again quite comparable to the
experimental result. The fitted curve for target-favoring search query volume
is again almost linear except steeper at the left end and the other two search
dimensions have ignorable impact, as shown in Figure
\ref{fig:Selection-bias-explained-case-123}(c).

\subsection{Empirical observations and discussions}

All three case studies above provide consistent empirical
evidence which validates the theory. First, a naive estimate of search ad ROAS
would lead to significant over-estimation. Second, a demand adjustment
helps reduce the bias but may be far from sufficient. Third,
the SBC method provides consistent selection bias correction and
its ROAS estimates are quite comparable to results from randomized
experimental studies.

\section{Case study with complex scenario}\label{sec:case4}

The advertiser in this case, called Case 4, had spend on more than a dozen different
media channels over the past three years, including both traditional
media and digital channels, with
search ads accounting for more than 1/3 of overall ads spend. The ads
spend and KPIs were collected on a daily basis. As in the above cases,
time series of search ad spend, search query volumes and sales all show strong
day-of-week patterns. The list of top 4 channels did not change over
the three years, which account for almost 90\% of overall ad spend.

The advertiser was never budget-constrained in the auction, so aside from
consumer demand, the two factors determining its search ad volume were its own
bidding (and related ad and page quality) and that of its competitors.
Thus we consider Figure \ref{fig:Causal-diagram-for-complex-2} as a
reasonable approximation to the true causal diagram.

\begin{figure}
\begin{centering}
\includegraphics[width=2.5in,height=2.5in]{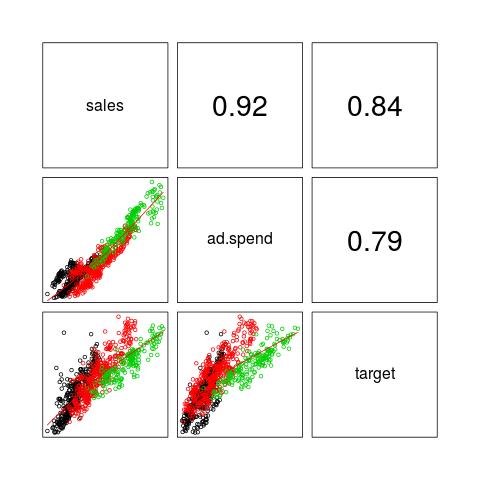}
\par\end{centering}

\caption{\label{fig:correlation-case-4}The correlation structure between daily search
  ad spend, target-favoring search volume and sales for Case 4
  (a complex scenario), where black, red and green dots represent the scatter
  plots (with scales removed) for 2013, 2014 and 2015 respectively.}
\end{figure}

Figure \ref{fig:correlation-case-4} shows the pairwise correlation
structure between sales, search ad spend and target-favoring search
volume, where the black, red and green colors mark the years of 2013,
2014 and 2015 respectively.
The pairwise scatterplots suggest somewhat different correlation between
target and (search ad spend, sales) over the three years. So we fit
the models for each year separately according to the additive form
\eqref{eq:fit-ssbc-simple}
and report the results in
Table \ref{tab:case-4}.

\begin{figure}
\begin{centering}
\includegraphics[width=5in,height=2in]{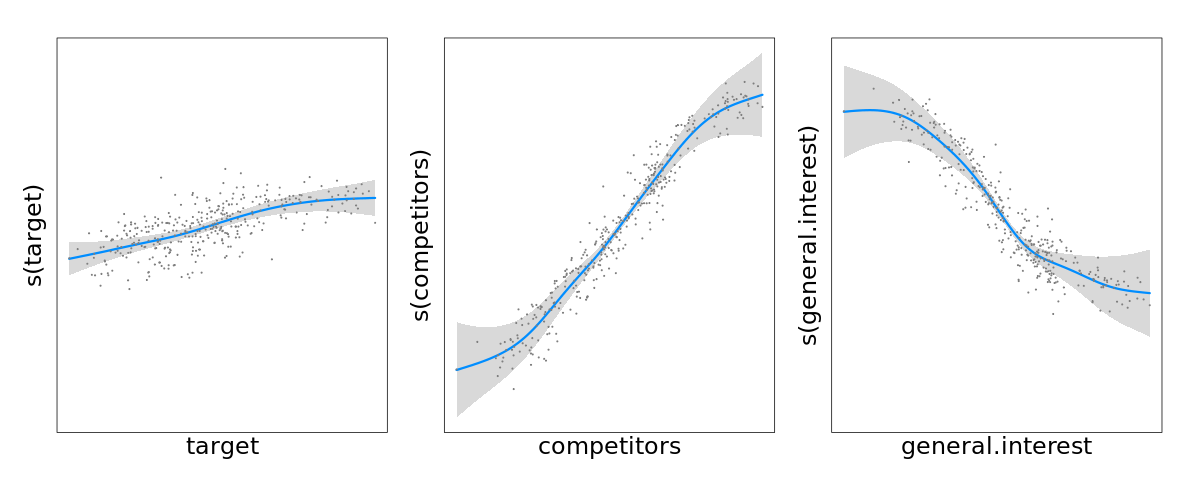}
\par\end{centering}

\caption{\label{fig:complex_case_gam}
  Selection bias explained by changes of target-favoring, competitors-favoring and general interest search query volumes in Case 4 for 2014; in each panel, x-axis represents query volumes and y-axis represents response values, where the
  response curves and 95\% confidence bands for the 3-dim search query volumes
  are fitted in an additive function as described in the regression
  \eqref{eq:fit-ssbc-simple}; the scatter plots are fitted function values
  plus model residuals.}
\end{figure}

The naive estimates of ROAS do not change much over the
years, while the SBC estimates keep growing and the estimate for
2015 is significantly higher than the estimate for 2013. This may
suggest that the advertising effectiveness has been improved gradually,
but we do not have randomized experimental results for reference. The response
curves for the search query volumes are reported in Figure
\ref{fig:complex_case_gam} for 2014 only, as they are similar for 2013 and
2015. Unlike previous cases, all three curves are statistically significant.

One might be curious why the response curve for the competitors-favoring search
volume is monotonically increasing as one would expect negative impact. It is
worth pointing out that the response curves for the 3-dim search query volumes
do not measure the causal impact of search volume on sales, but are the
projection of the sales due to consumer demand, organic search and other
non-search contributors onto the space of search queries, which serve the
role of bias correction for search ad.

Due to the relatively large sample size in this case, we
have also been able to fit the full regression model
\eqref{eq:fit-ssbc-simple-te}, with results comparable to the SBC results from the model \eqref{eq:fit-ssbc-simple}, as reported in Table \ref{tab:case-4}.\footnote{
 We have also performed the analysis on the data aggregated on the weekly basis,
and obtained higher estimates of absolute ROAS values for all methods,
probably due to search ad lag effect ignored by the daily-based models. The
effect of bias correction is similar.}

%%We have also fitted the above models for the same data but aggregated on the weekly basis. The point estimates from the Naive estimation, demand-adjusted and SBC methods are systematically somewhat higher. This may be due to lag effect, i.e. partial causal impact of search ads may not be materialized on the same day.

\begin{table}
\begin{centering}
\begin{tabular}{|c|c|c|c|c|}
\hline
 & Naive estimate & demand-adjusted  & SBC & SBC (full) \tabularnewline
\hline
%\hline
%2013  & 0.120 (.005)  & 0.073 (.014)  & 0.035 (.007) & 0.041 (0.008)\tabularnewline
%\hline
%2014  & 0.125 (.004)  & 0.058 (.009)  & 0.045 (.007) & 0.038 (0.007)\tabularnewline
%\hline
%2015  & 0.124 (.004)  & 0.106 (.004)  & 0.063 (.007) & 0.062 (0.007)\tabularnewline
%% R code:
%% x13 = c(0.120, 0.005, 0.073, 0.014, 0.035, 0.007, 0.041, 0.008)
%% x14 = c(0.125, 0.004, 0.058, 0.009, 0.045, 0.007, 0.038, 0.007)
%% x15 = c(0.124, 0.004, 0.106, 0.004, 0.063, 0.007, 0.062, 0.007)
%% round(c(x13, x14, x15) / 0.035, 2)
\hline
2013  & 3.43 (.14)  & 2.09 (.40)  & 1 (.20) & 1.17 (.23)\tabularnewline
\hline
2014  & 3.57 (.11)  & 1.66 (.26)  & 1.29 (.20) & 1.09 (.20)\tabularnewline
\hline
2015  & 3.54 (.11)  & 3.03 (.11)  & 1.80 (.20) & 1.77 (.20)\tabularnewline
\hline
\end{tabular}
\par\end{centering}
\vspace{0.1in}
\caption{Comparison of estimated ROAS for search ad in Case 4: Naive estimate,
	demand-adjusted estimate, and SBC for 2013, 2014 and 2015 respectively. Note that SBC (full) stands for results fitted from the SBC full regression model \eqref{eq:fit-ssbc-simple-te}, while SBC stands for results from the SBC model \eqref{eq:fit-ssbc-simple}. Here the SBC point estimate for 2013 is scaled to equal to one and all results and standard errors are indexed to that result.}
\label{tab:case-4}
\end{table}

\section{Discussion}\label{sec:discussion}

Measuring ad effectiveness with observational media mix data is hard. This
research focuses on search advertising and our major contributions are
as follows:

1) By looking into the causal diagrams of search
ads mechanism, we have derived a statistically principled method to estimate
search ad ROAS from MMM data for some common scenarios, where
search query data satisfy the back-door criterion for the causal effect
of paid search on sales.

2) Somewhat surprisingly,
for the scenarios identified by causal diagrams
in Figure \ref{fig:Causal-diagram-simple} and
\ref{fig:Causal-diagram-for-complex-2}, we have found that data on search ad,
relevant search queries and KPIs are sufficient to provide consistent
estimates of search ad ROAS, while data about non-search contributors
are not required. This is unlike traditional media mix models,
which usually fit a single regression with all relevant media and
control variables.

3) We have identified that one major assumption required by the theory
is satisfied when search ad spend is not constrained by its budget,
as is common practice in the industry.

4) Empirical studies on real cases in the simple scenario (causal diagram
in Figure \ref{fig:Causal-diagram-simple}) show promising results,
comparable to randomized geo experimental studies;
and an empirical study on a complex case scenario, without comparison to
randomized experimental studies, further shows significant difference between
the proposed SBC estimate and alternative estimates.

We have also validated the theory from various simulation studies
based on the simulator designed by \citet{zhang2017} recently,
where scenarios as depicted by causal diagrams in Figure
\ref{fig:Causal-diagram-simple}, \ref{fig:Causal-diagram-for-complex-1} and
\ref{fig:Causal-diagram-for-complex-2}
can be easily generated so that assumptions required by the causal
diagrams hold.

However, as in other observational studies, one must be always cautious in
interpreting the results as causal, because it is often hard to validate
the assumptions made by the causal diagrams.
We recommend MMM analysts to check with advertisers about the
assumptions; budget constraint or
budget planning across all media channels can help explain whether there is
any direct relationship between search ad spend and other relevant variables.
Below we list a few situations where we believe that
a straight-forward application of the proposed SBC estimate may be insufficient.

i) Data quality is poor. For example, top competitors are not identified
accurately and important search queries are missing from $V$.
As another example,
if ad impressions which did not lead to ad clicks had significant impact,
SBC which is currently based on search ad spend but ignores search ad
impressions, would under-estimate search ad ROAS. If the impact of search ad on
the KPI (e.g. store visits) is not immediate, i.e. there exists significant lag
effect, the estimate may be biased.

ii) It may be tempting to incorporate $V$ as an additional control variable
into traditional media mix models as described in \citet{jin2017}. This will
most likely reduce the coefficient of search ad, but the estimate may still be
biased.

iii) Existence of strong media mix synergy, where search ad impact may
heavily depend on simultaneous ad spends in other media channels.

iv) Existence of significant confounding effect from competitors' marketing
activities while competitors' information is not available.

v) The global marketing environment changes abruptly due to factors not
captured in the model, and search ad impact is affected correspondingly.

Nevertheless, by introducing Pearl's causal framework into media mix modeling,
our work provides a new research direction towards measuring media effect
truthfully in some practical scenarios. We expect to extend the research to
non-search media as well as to address some of the above issues in the future.

\section*{Acknowledgment}
We would like to thank Penny Chu, Nicolas Remy, Paul Liu,
Anthony Bertuca, Stephanie Zhang, Zhe Chen, Ling Leng, Katy Mitchell, Jon Vaver,
Tim Au, Shi Zhong, Xiaojing Huang, Conor Sontag, Patrick Hummel, Chengrui Huang,
Art Owen and Bob Bell for helpful discussion and support. Special thanks go to
Hal Varian and Tony Fagan for many insightful discussions and review comments to
improve the paper quality.
The work was partially motivated by Professor Peter Rossi's keynote speech
at Google's MMM summit in NYC in January 2016.

\section*{Appendix}

\subsection{A more realistic search causal diagram}

\begin{figure}
\begin{centering}
\includegraphics[width=2.5in,height=2.5in]{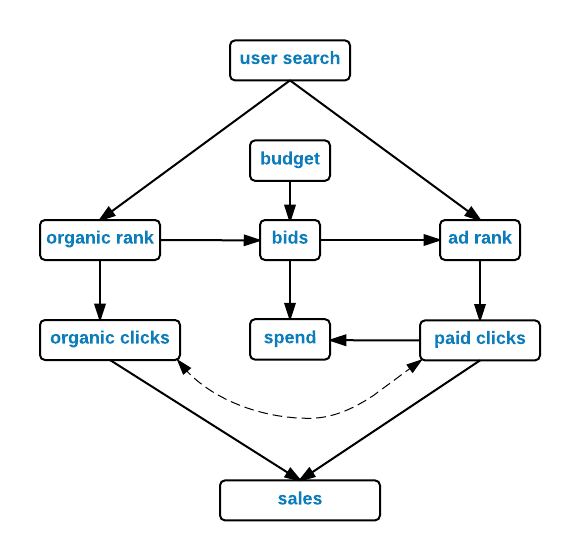}
\par\end{centering}
\caption{\label{fig:hal_search_diagram}A more precise causal diagram for search
  ad at a query level, where the dashed edge between organic click and paid
  click represents potential cannibalization effect.}
\end{figure}

Instead of Figure \ref{fig:A-causal-diagram}, a more precise causal diagram for
search ad can be described by Figure \ref{fig:hal_search_diagram}.\footnote{This
  diagram was shared by Hal Varian.}
Since predicted click-through rate is part of the auction scores, there is a
directed edge from paid clicks to ad rank but at a later time, not shown on the
diagram for simplicity.
The diagram suggests that:
1) bids and budgets are the causes while ad spend and ad clicks are the
intermediate outcomes in the MMM problem, therefore, measuring
the effect of ad spend on sales may be an ill-posed problem;
2) Organic rank and organic clicks may be a confounding factor.
One could also imagine paid clicks causing organic search.
The more it is advertised, the more people recognize the brand and the more
they search for the brand. So the ad may have impression value that stimulates
searches, but the effect may be weaker.

Instead of using ad spend, a better formulation can be
made w.r.t. ad impressions as described in Figure \ref{fig:A-causal-diagram}.
Nevertheless, our case studies suggest that one may still obtain reasonable
estimates under some common scenarios.

We have not studied how to incorporate organic rank into the model because
organic rank is often stable during a short time window. However, it can
be used to further improve dimension reduction of relevant search queries.

One must be cautious in consideration of organic clicks as a confounding factor.
As a toy example, suppose user searches do not change and nothing else changes
except that ads grow. Assume no lag effect and organic rank is stable.
Then the effect of ads change (e.g. changing bids or budgets) can be measured
simply by the change in sales.
Since organic clicks decrease due to negative correlation with ad clicks, i.e.
cannibalization effect (see \citep{hal2009adclick,blake2015} for real examples), bringing organic clicks into the model would bias
the estimate. In fact, self-loops are not supported in Pearl's causal diagram.
To break the loop in Figure \ref{fig:hal_search_diagram}, it looks more
reasonable to use the direction "paid clicks $\rightarrow$ organic clicks",
instead of the opposite, and then organic clicks should not be controlled
according to the back-door criteria.

\subsection{More examples where Figure \eqref{fig:Causal-diagram-simple} does not hold} \label{subsec:counter-examples}

We provide a few more examples below where condition (b) is violated and the causal diagram identified by Figure \eqref{fig:Causal-diagram-simple} does not hold.

Example 1. A movie may have just won a prestigious award. This could have the
effect of increasing both consumer demand (i.e. search queries for the movie)
and click-through rates on search ads for the movie. Then there can be a direct
edge from consumer demand to $X$ which does not go through search queries.

Example 2. Assuming auction factors stay constant, any situation which affects
both consumer demand and click-through rates, can lead to a direct edge from
consumer demand to $X$. The Equifax data breach\footnote{https://www.consumer.ftc.gov/blog/2017/09/equifax-data-breach-what-do} is one such example, which can
cause a loss of confidence in the advertiser, leading to much lower CTRs and hence lower $X$.

Example 3. An advertise increases its search ad bids and also reduces its
product price due to factors in its business that have no effect on overall
consumer demand, such as a reduction in cost-of-goods. The advertiser's sales
and search ad volume will go up, but the effect of the search ads on sales is
confounded by its price change and is not identified.

Example 4. An advertiser's competitor increases its search ad bid and also
reduces its product price due to factors in its business that have no effect
on overall consumer demand, such as a reduction in cost-of-goods. The
advertiser's sales and search ad volume will go down, but the effect of the
search ads on sales is confounded by competitor price changes and is not
identified.

\printbibliography
\end{document}